\documentclass[aps,showpacs,amsmath,amssymb,superscriptaddress]{revtex4}

\usepackage{graphicx}
\usepackage{amssymb,amsmath}
\usepackage{dcolumn}
\usepackage{bm}

\begin{document}
\title{Photon emission induced by elastic exciton--carrier scattering in semiconductor quantum wells}

\author{H. Ouerdane}
\affiliation{LASMEA, CNRS-Universit\'e Blaise Pascal, 24 Avenue des Landais, 63177 Aubi\`ere Cedex France}
\affiliation{Physics, School of Engineering and Physical Sciences, Heriot-Watt University, Edinburgh EH14 4AS, UK}
\author{R. Varache}
\affiliation{LASMEA, CNRS-Universit\'e Blaise Pascal, 24 Avenue des Landais, 63177 Aubi\`ere Cedex France}
\author{M. E. Portnoi}
\affiliation{School of Physics, University of Exeter, Exeter EX4 4QL, UK}
\author{I. Galbraith}
\affiliation{Physics, School of Engineering and Physical Sciences, Heriot-Watt University, Edinburgh EH14 4AS, UK}

\begin{abstract}
We present a study of the elastic exciton--electron ($X-e^-$) and exciton--hole ($X-h$) scattering processes in semiconductor quantum wells, including fermion exchange effects. The balance between the exciton and the free carrier populations within the electron-hole plasma is discussed in terms of ionization degree in the nondegenerate regime. Assuming a two-dimensional Coulomb potential statically screened by the free carrier gas, we apply the variable phase method to obtain the excitonic wavefunctions, which we use to calculate the 1$s$ exciton--free carrier matrix elements that describe the scattering of excitons into the light cone where they can radiatively recombine. The photon emission rates due to the carrier-assisted exciton recombination in semiconductor quantum-wells (QWs) at room temperature and in a low density regime are obtained from Fermi's golden rule, and studied for mid-gap and wide-gap materials. The quantitative comparison of the direct and exchange terms of the scattering matrix elements shows that fermion exchange is the dominant mechanism of the exciton--carrier scattering process. This is confirmed by our analysis of the rates of photon emission induced by electron-assisted and hole-assisted exciton recombinations.
\end{abstract}


\maketitle

\section{Introduction}
\label{intro}
Scattering processes involving excitons are at the heart of a wide range of phenomena in semiconductor optics. Excitons are often treated as elementary bosons, but the complexity of the scattering problem obviously lies in their composite nature: excitons are Coulomb-correlated quasiparticles made of two fermions, a conduction band electron and a valence band hole. It is only recently that a many-body theory of exciton scattering was put on firm grounds in a series of papers by Combescot \emph{et al} (see e.g. Refs.~\cite{COM04,COM07} and references therein). Their formalism, based on fermion commutation techniques, allows one to calculate correctly the scattering matrix elements as well as the transition rates of two excitons. In the present paper, we are only concerned with the simpler problem of exciton--carrier scattering, i.e. a 3-body problem with well defined interaction potentials between the three scattering partners, including exchange effects explicitly. More precisely, we are interested in exciton--free carrier scattering in III--V GaAs-based and II--VI ZnSe-based quantum-well systems.

Early investigations of the microscopic mechanisms yielding low lasing thresholds and high optical gain in bulk semiconductor lasers included detailed theoretical and experimental studies of radiative recombination involving exciton scattering. Various scattering processes including $X-X$, $X-$ LO-phonon, $X-e^-$ scattering \cite{BEN69,HAU77,KOC78} and $X-h$ scattering \cite{HAU77,KOC78} were studied and compared, and it was reported that $X-e^-$ scattering is a process yielding important optical gain in bulk materials \cite{BEN69,HAU77,KOC78}. In particular, Haug and co-workers gave a detailed study of the temperature dependence of the lasing thresholds for each of the scattering process \cite{HAU77,KOC78}. Later, Feng and Spector studied the exciton--free carrier elastic and inelastic scatterings in quantum wells to compute the related cross sections and exciton linewidths, but they did not account for carrier exchange in their analysis \cite{FEN87}. Here, the aim of our work is two-fold: first we want to compare the respective contributions of the direct and the exchange terms to the scattering matrix elements; then we want to study quantitatively the exchange term dependence on the carrier effective mass and its impact on carrier-assisted exciton radiative recombination in quantum wells.

The interest for extensive studies of photoluminescence (PL) spectra \cite{HEN80,SCH86,KUS89,CHR90,COL91,BLO91,SHA92,BLO93,GAL96,NUS97,KIR98,KIR99,GAL05,CHA04,SZC05,NAK06} lies in the fact that they are useful to study a rich variety of phenomena in semiconductor physics and allow non-destructive characterization of semiconductors. In the case of mixed exciton/electron-hole plasmas, many processes occur that contribute to PL spectra. These processes include: direct exciton recombination, electron-hole recombination and several scattering processes involving quasi-particles such as excitons, phonons and excited carriers. The importance of each contribution to PL depends on the thermodynamic properties of the plasma and hence on the excitation conditions as well as the type of material under consideration that determine the populations and temperatures of each quasi-particle gas. Moreover, the dimensionality of the system is also of great importance since quantum confinement not only enhances the effective strength of the Coulomb interaction between the carriers, but changes drastically the density of states of the particles. 

Early works by Kubo, Martin and Schwinger (KMS) \cite{KUB57,MAR59} suggest that in quasi-equilibrium the intensity of PL is proportional to the absorption coefficient times the Bose distribution describing the photon gas interacting with the semiconductor medium. Detailed studies of the excitonic resonance features and their dynamics through PL spectra based on the KMS approach led to the interpretation of the build-up of the excitonic resonance below the gap as direct evidence of excitonic formation on the sub-nanosecond timescale, see e.g. Refs.~\cite{KUS89,BLO93,NUS97}. However, calculations by Kira \emph{et al} \cite{KIR98,KIR99} clearly show that there is no straightforward connection between exciton formation and the presence of an excitonic peak in the emission: free plasma emission can occur at the exciton energy. Galbraith \emph{et al} \cite{GAL05} reported later on the existence of luminescence at the exciton energy in GaAs-based multiple quantum well, consistent with the theoretical work of Kira \emph{et al} \cite{KIR98,KIR99}, and Chatterjee \emph{et al} \cite{CHA04} identified conditions under which the PL emission at the exciton resonance after excitation in the continuum is dominated by the free electron-hole plasma or by an incoherent exciton population at low temperature (but they could not determine experimentally the fraction of excitons contributing to PL). This contradicts the findings reported by Szczytko \emph{et al} \cite{SZC05} who claimed that excitons provide the dominant contribution to the luminescence signals at the exciton energy, and concluded that for densities, temperatures, and time scales actually used in time-resolved experiments the Coulomb correlated plasma contribution may be neglected. The conclusion on the role of a finite exciton population based on the results obtained from simple rate equations used by Szczytko \emph{et al} may require further investigation to be confirmed, but models that take into account the existence of a finite population of excitons in excited QWs deserve nonetheless attention.

In our model, we assume that the exciton and carrier populations at room temperature in a low density regime have reached thermodynamic equilibrium. During the course of the scattering process, an exciton leaves its initial state and reaches the photon line before radiative recombination, transferring its momentum to its scattering partner, a free carrier. The exciton--electron scattering was found to be an important mechanism producing photoluminescence in bulk materials \cite{BEN69,HAU77,KOC78,GAL96} and optical gain in thin films \cite{NAK06}. It is thus of interest to investigate this mechanism in mid-gap and wide-gap quantum wells, and also obtain further insight by including the exciton--hole scattering in the analysis. Therefore, the object of the present work is \emph{not} the calculation of the full PL spectra for which a full microscopic theory \cite{KIR98,KIR99,CHA04} is required, but the quantitative study of the relative importance of $X-e^-$ and $X-h$ scattering mechanisms, including exchange. Direct exciton recombination, known to yield a strong emission at exciton energy \cite{CHR90,COL91}, as well as scattering processes other than that of excitons with free carriers are beyond the scope of the present work and hence are not included in the paper.

A complete treatment of the balance between bound excitonic states, scattering states and free carriers is a difficult problem that has attracted attention for several years \cite{KRA75,ZIM85,KRA86,ZIM88,POR98,POR99,NIK04}. In the case of QWs, advances were made in Ref.~\cite{POR99} where the degree of ionization, $\alpha$, of a non-degenerate two-dimensional (2D) electron-hole plasma was calculated as the ratio $n^0/n$ of the density of free carriers, $n^0$, to the total plasma density, $n=n^0+n^{\rm {corr}}$ where $n^{\rm {corr}}$ is the density of Coulomb correlated carriers \cite{ZIM85}. To account for occupation of scattering states and overlap of exciton wavefunctions at moderate and high densities this concept was further refined in Ref.~\cite{NIK04}, so that the definition of $\alpha$ is given by $\alpha = 1 - n_{{\rm ab}}/n_{\rm a}$, where $n_{\rm a}$ is the total density of carriers of type ${\rm a}$ and $n_{{\rm ab}}$ originates from the interaction of carriers of type ${\rm a}$ with carriers of type ${\rm b}$ (${\rm a}\neq{\rm b}$). The ionization degree is now formulated in terms of elementary excitations (electrons and holes). Note that in the low density limit, $n_{{\rm ab}}$ can be identified to the exciton density. The degree of ionization whose values are between 0 and 1, is both density and temperature dependent. Authors of Ref.~\cite{POR99} demonstrated that for wide-gap semiconductor quantum-wells at room temperature, the equilibrium consists of an almost equal mixture of correlated electron-hole pairs and uncorrelated free carriers whereas this is not the case for mid-gap materials.

For a plasma temperature of 300 K, we ensure the occupation density is small enough (nondegenerate limit) to neglect phase-space filling effects. Under these conditions, the scattering between excitons and carriers involves mostly thermalized 1$s$ excitons with a finite center of mass momentum and free carriers. This assumption is valid at room temperature and moderate densities where the 1$s$ exciton population dominates all other Coulomb correlations in the electron-hole plasma. Calculations in an unscreened Coulomb potential with hydrogen atom-like exciton wavefunctions have been reported by Kavokin \cite{KAV03}. In the present work, the electron-hole bound states are computed assuming a statically screened Coulomb potential (whose asymptotic behavior permits the application of the variable phase method \cite{CAL67}).

In Section II we detail the basic assumptions made for the model of the interacting 2D electron-hole plasma and define the ionization degree. Section III is devoted to the derivation of the exciton wavefunctions from the variable phase theory. In Section IV we calculate the exciton--free carrier scattering matrix elements that we include into Fermi's golden rule to compute the related emission rates. Numerical results are presented and discussed in Section V, where photon emission rates induced by exciton--carrier scattering in ZnSe and GaAs QWs are compared.

\section{Ingredients of the model}

\subsection{Underlying assumptions}

In this work we  assume  the quantum-well structures to be ideal two-dimensional systems in a two-band model with parabolic dispersion, neglecting detail related to valence-band mixing. We consider a neutral low density, 2D plasma composed of interacting electrons and holes, at room temperature. We thereby remain in the Boltzmann regime defined by \cite{HUA87}:

\begin{equation} \label{eq1}
	n\lambda_{m_{\rm c}}^2/g \ll 1,
\end{equation}

\noindent where $n$ is the 2D carrier density, $\lambda_{m_{\rm c}}=(2\pi\hbar^2/m_{\rm c}k_{\rm B}T)^{1/2}$ the thermal wavelength and $g$ the spin degeneracy of the confined carriers of effective mass $m_{\rm c}$. For $T$ = 300 K, Eq.~(\ref{eq1}) is satisfied for $n \ll 1.7 \times 10^{12}$ cm$^2$ for ZnSe-based QWs and $n \ll 7.2 \times 10^{11}$ cm$^2$ for GaAs-based QWs \cite{POR99}.

Due to spatial confinement of the carriers in QWs the Coulombic interaction between electrons and holes  is  enhanced. However, even at moderate plasma densities carrier-carrier interactions are weakened because of the density-dependent screening. Screening is one of the most important manifestations of the complex many-body interaction in the electron-hole plasma and the simplest approach is the use of the 2D statically screened potential \cite{STE67}:

\begin{equation} \label{eq2}
  V_{\rm s}(\rho) = \frac{\displaystyle e^2}{\displaystyle \epsilon}\int_{0}^{\infty} \frac{\displaystyle qJ_0(q\rho)}{\displaystyle q+q_{\rm s}}~{\rm d}q,
\end{equation}

\noindent where $J_0$ is the Bessel function; $q_{\rm s}$, in the Boltzmann limit, is known as the 2D Debye-H\"uckel screening wavenumber (which depends on temperature and carrier density), $\epsilon$ is the static dielectric constant of the semiconductor and $\rho$ the inter-carrier distance. Equation (\ref{eq2}) describes the electron-electron and hole-hole repulsion; the attractive electron-hole potential is obtained by changing the overall sign.

The above integral may also be written in terms of special functions:

\begin{equation} \label{eq2b}
  V_{\rm s}(\rho) = \frac{\displaystyle e^2}{\displaystyle \epsilon} \left[ \frac{\displaystyle 1}{\displaystyle \rho}-\frac{\displaystyle \pi}{\displaystyle 2}~ \left( {\bf H}_0(q_{\rm s}\rho)-N_0(q_{\rm s}\rho)\right)\right],
 \end{equation}
 
\noindent where $N_0$ and ${\bf H}_0$ are the Neumann and Struve functions respectively \cite{ABR72,GRA80}.

The total screening is the sum of electron and hole plasma screenings. As such we neglect the weak screening by neutral excitons. 

\subsection{The degree of ionization}

In QWs where Coulomb forces are enhanced because of the spatial confinement of the electrons and holes, we expect a finite population of Coulomb correlated quasi-particles in the plasma even at room temperature. The balance between those correlations and the free carriers is obtained by calculating the degree of ionization, $\alpha$ \cite{NIK04}:

\begin {equation} \label{eq3}
  \alpha = 1 - \frac{\displaystyle n_{\rm eh}}{\displaystyle n_{\rm e}} = 1 - \frac{\displaystyle n_{\rm eh}}{\displaystyle n_{\rm h}}.
\end {equation}

\noindent In the above equation, the total density of a carrier of type ${\rm a}$ is defined as \cite{NIK04}:

\begin {equation} \label{eq4}
	n_{\rm a} = n_{\rm a}^0 + n_{\rm aa} + n_{\rm ab},
\end {equation}

\noindent where $n_{\rm a}^0$ is the density of free carriers of type ${\rm a}$ and $n_{\rm aa}$ originates from the interaction between carriers of the same type (${\rm ee}$ or ${\rm hh}$). When $\alpha$ is close to unity, the thermodynamic properties of the electron-hole plasma are those of the ideal gas (defined by $\alpha$ = 1). For lower values of the degree of ionization the thermodynamic properties of the plasma deviate from those of the ideal gas and Coulomb correlations have to be considered.

Detailed calculations of the degree of ionization and further discussion on the statistical mechanics of the 2D electron-hole plasma can be found in Refs.~\cite{POR98,POR99,NIK04}. Here we only stress that the degree of ionization is obtained from the calculation of the partition function of the 2D electron-hole plasma. Scattering state contributions to the partition function have to be considered in addition to the bound-state sum as a proper account of scattering eliminates singularities in thermodynamic properties of the non-ideal 2D gas caused by the emergence of additional bound states as the strength of the attractive potential is increased. Inclusion of the scattering states also leads to a strong deviation from the standard law of mass action. Additionally, note that exchange between carriers within the electron-hole plasma is taken into account in the calculation of the electron-electron and hole-hole parts of the partition function \cite{POR99}.

\begin {figure}[!rh]
\centering
\scalebox{.33}{\rotatebox{0}{\includegraphics*{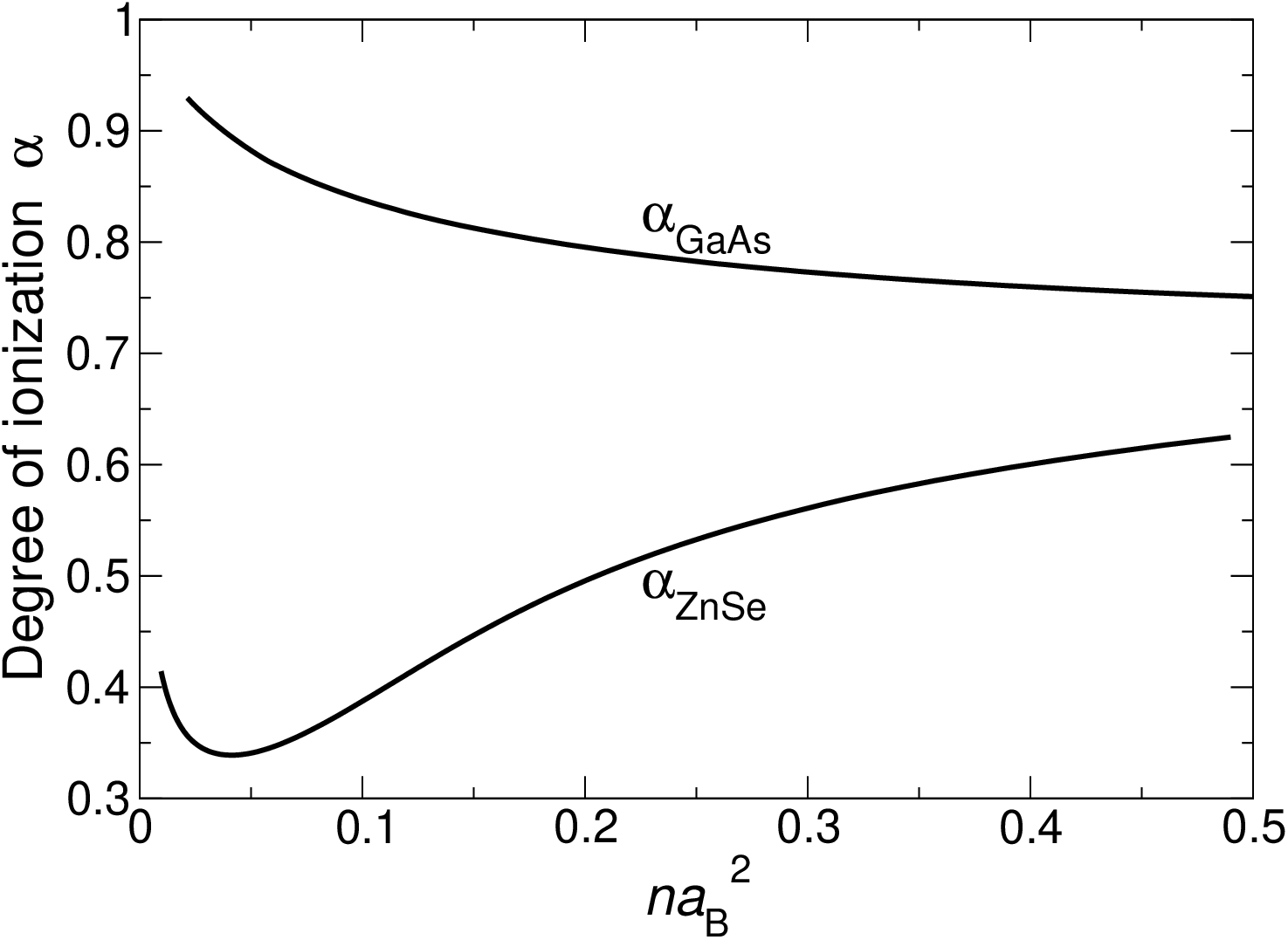}}}
\caption{Ionization degree evaluated for GaAs and ZnSe quantum wells in the Boltzmann regime for $T = 300$ K, as a function of the 2D plasma density scaled to the square of the excitonic Bohr radius, $a_{\rm B}$.}\label{fig1}
\end {figure}

The ionization degrees, $\alpha_{\rm{GaAs}}$ and $\alpha_{\rm {ZnSe}}$, are shown as functions of plasma density $n$ on Fig.~\ref{fig1} for GaAs and ZnSe quantum-wells at temperature $T=300$ K. In the nondegenerate regime, the function $\alpha_{\rm{GaAs}}(n)$ is monotonic whereas $\alpha_{\rm {ZnSe}}(n)$ exhibits a minimum at low density. The presence of the minimum is due to the influence of Coulomb screening on the density dependence of the ionization degree: at very low density screening is negligible and $n_{\rm a}^{\rm {corr}}$ is proportional to the square of the total plasma density so $\alpha$ is a decreasing function of $n$; however with $n$ increasing further, plasma screening becomes more and more important and hence reduces the strength of Coulomb correlations so $n_{\rm a}^{\rm {corr}}$ is a decreasing function of $n$ and $\alpha$ increases. This density dependence is more dramatic in ZnSe QWs where Coulomb forces are much stronger than they are in GaAs QWs.  Note that $\alpha_{\rm{GaAs}}(n)$ and $\alpha_{\rm {ZnSe}}(n)$ not only differ qualitatively but also quantitatively: in GaAs QWs the electron-hole plasma is dominated by free carriers whereas it contains comparable densities of correlations and free carriers in ZnSe QWs. Knowledge of the degree of ionization is of great importance as the emission mechanisms depend on the nature of the plasma. For further detail see Ref.~\cite{POR99}.

\section{Two-dimensional exciton wavefunction}

In this section we calculate the wavefunction of an electron-hole bound state in the statically screened potential, Eq. (\ref{eq2}). We suppose that the interaction between the two particles depends only on the relative distance $\rho = |{\bm r}_{\rm e}-{\bm r}_{\rm h}|$, and we split the problem into two parts: the study of the relative motion of the two particles and the study of the motion of the center of mass which does not depend on the interaction. The total Hamiltonian can then be written as: ${\hat H_{\rm {tot}}}={\hat H_{\rm {cm}}}+{\hat H_{\rm {rel}}}$. To study the relative motion problem we apply the variable phase method of scattering theory, neglecting the intrinsic spin effects.

The Schr\"odinger equation for the radial wavefunction of the relative motion has the following form:

\begin {equation} \label{eq11}
	\left[\frac{\displaystyle {\rm d}^2}{\displaystyle {\rm d}\rho^2}~ +\frac{\displaystyle 1}{\displaystyle \rho}\frac{\displaystyle {\rm d}}{\displaystyle {\rm d}\rho}~+\kappa^2-U(\rho)-\frac{\displaystyle m^2}{\displaystyle \rho^2}\right]~ R_{m,\kappa}(\rho) = 0,
\end {equation}

\noindent for a given value of $m$ , the azimuthal quantum number and where $k^2=2m_{\rm r}E/\hbar^2$ and $U(\rho)=2m_{\rm r}V_{\rm s}(\rho)/\hbar^2$. 

For the bound states, the energy $E$ is negative and we introduce the imaginary wavenumber $k=i\kappa$. As the potential vanishes at large distances, the solution of the radial equation Eq.~(\ref{eq9}) can be approximated for large $\rho$ by the solution of the free Bessel equation, which is a linear combination of the modified Bessel functions of the first and second kind. Then, the solution of the radial Schr\"odinger, Eq.~(\ref{eq9}), can be written as follows \cite{POR99}:

\begin {equation} \label{eq12}
	R_{m,\kappa}(\rho) = A_m~ \left( I_m(\kappa\rho) \cos\eta_m + \frac{\displaystyle 2}{\displaystyle \pi}~K_m(\kappa\rho) \sin\eta_m \right),
\end {equation}

\noindent where $I_m(\kappa\rho)$ and $K_m(\kappa\rho)$ are the modified Bessel functions of the first and second kinds respectively while the phase shift $\eta_m$ characterises their admixture and $A_m$ is the wavefunction amplitude.

To solve the problem for all $\rho$, not just $\rho \rightarrow \infty$, the  phase shift $\eta_m$ and the amplitude $A_m$ are both considered not as constants but as explicit functions of $\rho$ and $\kappa$ in the 2D formulation of the variable phase method \cite{CAL67}. Following Ref.~\cite{POR99}, we insert Eq.~(\ref{eq12}) into Eq.~(\ref{eq11}) and find that the phase shift satisfies the following first order, non-linear differential equation of the Ricatti type:

\begin{equation}\label{eq13}
\frac{\displaystyle {\rm d}}{\displaystyle {\rm d}\rho}~ \eta_{m,\kappa}(\rho) = -\frac{\displaystyle \pi}{\displaystyle 2}~\rho~ U(\rho)\times\left(I_m(\kappa\rho)\cos \eta_{m,\kappa}(\rho) + \frac{\displaystyle 2}{\displaystyle \pi}~ K_m(\kappa\rho) \sin \eta_{m,\kappa}(\rho)\right)^2.\nonumber
\end{equation}

\noindent Equation~(\ref{eq13}) is called the phase equation and should be solved with the boundary condition:  $\eta_{m,\kappa}(0) = 0$, thus ensuring that the radial function does not diverge at $\rho=0$. For the bound states the diverging solution vanishes, thus implying the asymptotic condition:

\begin {equation} \label{eq17}
	\lim_{\rho\rightarrow\infty}\eta_{m,\kappa}(\rho) = (\nu-1/2)\pi,
\end {equation}

\noindent where $\nu$ enumerates the bound states for a given $m$. The number of non-zero nodes of the radial wave function is given by $\nu -1$.

Similarly the amplitude $A_{m,\kappa}(\rho)$ satisfies the following equation:

\begin{equation} \label{eq15}
\frac{\displaystyle {\rm d}}{\displaystyle {\rm d}\rho}~A_{m,\kappa}(\rho) = 
A_{m,\kappa}(\rho)~
\frac{\displaystyle I_m(\kappa\rho)\sin\eta_{m,\kappa}(\rho) - \frac{\displaystyle 2}{\displaystyle \pi}~ K_m(\kappa\rho)\cos\eta_{m,\kappa}(\rho)}{\displaystyle I_m(\kappa\rho)\cos\eta_{m,\kappa}(\rho) + \frac{\displaystyle 2}{\displaystyle \pi}~ K_m(\kappa\rho)\sin\eta_{m,\kappa}(\rho)}\times~
\frac{\displaystyle {\rm d}}{\displaystyle {\rm d}\rho}~\eta_{m,\kappa}(\rho),
\end{equation}

\noindent which is coupled to the phase equation, Eq.~(\ref{eq13}), whose solution is computed first.

The total bound exciton wavefunction is hence given by the product of the relative wavefunction

\begin{equation}\label{eq16}
\psi_{m,\kappa}(\rho,\varphi) = \left(I_m(\kappa\rho) \cos\eta_{m,\kappa}(\rho)+
\frac{\displaystyle 2}{\displaystyle \pi}~ K_m(\kappa\rho)\sin\eta_{m,\kappa}(\rho)\right)\times A_{m,\kappa}(\rho)e^{im\varphi},
\end{equation}

\noindent and the exciton center of mass motion wavefunction characterized by the plane wave $\phi_{{\bm k}_{\rm {cm}}}({\bm R}) = \exp (-i{\bm k}_{\rm {cm}}\cdot{\bm R})/\sqrt{{\mathcal A}}$, where ${\bm R}=(m_{\rm e}{\bm r}_{\rm e}+m_{\rm h}{\bm r}_{\rm h})/(m_{\rm e}+m_{\rm h})$ is the exciton center of mass coordinate, ${\bm r}_{\rm e}$ and ${\bm r}_{\rm h}$ the carrier coordinates , ${\mathcal A}$ is the surface area of the 2D system and $k_{\rm {cm}}={\sqrt {2ME_{\rm {cm}}}}/\hbar$ the exciton center of mass momentum. Calculation of the scattering states is detailed in Ref.~\cite{POR99}.

\section{Photon emission induced by exciton-carrier scattering}

\subsection{The scattering matrix elements}

In this section we are concerned with the scattering of 1$s$ excitons with free carriers: the excitons transfer their momenta and energies to the free carriers and reach the photon line where they recombine and thus emit a  photon. Two possible spin configurations should be considered: depending on its total spin state an exciton may or may not radiatively recombine because of selection rules. In the $X-e^-$ scattering process, the triplet state is defined with parallel electron spins and the singlet state with antiparallel electron spins (the same obviously applies for hole spins in the case of $X-h$ scattering). If, for instance, the exciton spin is $+1$, the exciton is said to be bright and if it is $+2$, the exciton is dark. Therefore, exchange between the free carrier and the bound carrier changes the nature of the exciton in the singlet configuration, while the total exciton spin is conserved in the triplet configuration. 

The carriers interact via the screened potentials defined in Eq.~(\ref{eq2}) and the matrix element $V_{\rm {scat}}^{\rm {cX}}$, ${\rm c} \equiv {\rm e},{\rm h}$, can be written as a sum (difference) of a direct term $V_{\rm {dir}}^{\rm {cX}}$ and an exchange term  $V_{\rm {exch}}^{\rm {cX}}$ in the triplet (singlet) configuration:

\begin{equation} \label{eq19}
V_{\rm {scat}}^{\rm {cX}}({\bm k}_{\rm {cm}},{\bm k}_2) = V_{\rm {dir}}^{\rm {cX}}(k_{\rm {cm}}) \pm V_{\rm {exch}}^{\rm {cX}}({\bm k}_{\rm {cm}},{\bm k}_2).
\end{equation}

\noindent As shown in Appendix A, in the case of exciton--electron scattering:

\begin{equation} \label{eq20}
V_{\rm {dir}}^{\rm {eX}} = \frac{\displaystyle e^2}{\displaystyle 2\epsilon_0\epsilon_{\rm r}} \frac{\displaystyle {\widetilde {R_0^2}}_{,\kappa}(\gamma_{\rm h}k_{\rm {cm}}) - {\widetilde {R_0^2}}_{,\kappa}(\gamma_{\rm e}k_{\rm {cm}})}{\displaystyle k_{\rm {cm}}+q_{\rm s}},
\end{equation}

\noindent and

\begin{eqnarray} \label{eq20b}
V_{\rm {exch}}^{\rm {eX}}({\bm k}_{\rm {cm}},{\bm k}_2) & = & \frac{\displaystyle e^2}{\displaystyle 2\epsilon_0\epsilon_{\rm r}} \left( \int \frac{\displaystyle \widetilde{R}_{0,\kappa}(-{\bm k}_2 - {\bm q})\widetilde{R}_{0,\kappa}({\bm k}_2+\gamma_{\rm h}{\bm k}_{\rm {cm}}+{\bm q})}{\displaystyle q+q_{\rm s}}~{\rm d}{\bm q}+ \widetilde{R}_{0,\kappa}({\bm k}_2 +\gamma_{\rm h}{\bm k}_{\rm {cm}}) \int \frac{\displaystyle \widetilde{R}_{0,\kappa}(-{\bm k}_2 - {\bm q})}{\displaystyle q+q_{\rm s}}~{\rm d}{\bm q} \right.
\nonumber\\
&& - \left. \widetilde{R}_{0,\kappa}({\bm k}_2) \int \frac{\displaystyle \widetilde{R}_{0,\kappa}({\bm k}_2+\gamma_{\rm h}{\bm k}_{\rm {cm}}-{\bm q})}{\displaystyle q+q_{\rm s}}~{\rm d}{\bm q}\right),
\end{eqnarray}

\noindent where $\gamma_{\rm c} = m_{\rm c}/M$. Note that the minus sign before the third term in the equation above may change the overall sign of the exchange term when it is computed as a function of $k_{\rm {cm}}$. To obtain these two terms for the exciton--hole scattering, one should simply swap the effective masses $m_{\rm e}$ and $m_{\rm h}$ where appropriate and change the overall sign.

In Eqs.~(\ref{eq20})and (\ref{eq20b}), the quantities $\widetilde{R_0}$ and $\widetilde{R_0^2}$ respectively denote the Fourier transforms of the $m=0$ radial wavefunction in Eq.~(\ref{eq16}) and its square. Since the free carriers wavefunctions are taken as plane waves, the direct term $V_{\rm {dir}}^{\rm {cX}}$, c=e,h, is independent of the scattering free carrier momentum ${\bm k}_2$ and  only depends on the energy of the bound state characterized by the wavenumber $\kappa$, the effective masses of the carriers $m_{\rm e}$ and $m_{\rm h}$, and the initial exciton momentum $k_{\rm {cm}}$. We also find that $V_{\rm {dir}}^{\rm {hX}}(k_{\rm {cm}}) = - V_{\rm {dir}}^{\rm {eX}}(k_{\rm {cm}})$.

\subsection{Emission rates}

In this section we give the expressions of the exciton--carrier scattering contributions to the photon emission rates, $R_{\rm {eX}}(\hbar\Omega)$ and $R_{\rm {hX}}(\hbar\Omega)$. The expression for $R_{\rm {eX}}(\hbar\Omega)$, derived from Fermi's golden rule, is taken from Ref.~\cite{BEN69}:

\begin {equation} \label{eq7}
R_{\rm {eX}}(\hbar\Omega) = \sum_{{\bm k}_{\rm {cm}}}\sum_{{\bm k}_2} C_{{\bm k}_{\rm {cm}},{\bm k}_2}N_{{\bm k}_{\rm {cm}}}N_{{\bm k}_2},
\end {equation}

\noindent with

\begin{equation} \label{eq8}
C_{{\bm k}_{\rm {cm}},{\bm k}_2} = \frac{\displaystyle 2\pi}{\displaystyle \hbar}~|V_{\rm {scat}}|^2  \left( \frac{\displaystyle 4\pi\xi \hbar\Omega /E_{\rm g}}{\displaystyle ( 1-\hbar^2\Omega^2 /E_{\rm g}^2)^2 +4\pi\xi}\right)\times~ \delta \left (E_{\rm X}-\hbar\Omega-\frac{\displaystyle \hbar^2}{\displaystyle 2m_{\rm e}}(k_{\rm {cm}}^2+2{\bm k}_{\rm {cm}} \cdot {\bm k}_2)\right),
\end{equation}

\noindent where $V_{\rm {scat}}(k_{\rm {cm}},k_2)$ are the exciton-carrier scattering matrix elements given by Eqs.~(\ref{eq20}) and (\ref{eq20b}), ${\bm k}_{\rm {cm}}$ the exciton center of mass wavevector, and ${\bm k}_2$ the free electron wavevector. The exciton and free carrier distributions, $N_{{\bm k}_{\rm {cm}}}$ and $N_{{\bm k}_2}$, are:

\begin {equation} \label{eq9}
   N_{{\bm k}_{\rm {cm}}} = \frac{\displaystyle 2\pi\beta\hbar^2}{\displaystyle M}~ (1-\alpha)n~\exp \left(\!-\beta~ \frac{\displaystyle \hbar^2k_{\rm {cm}}^2}{\displaystyle 2M}\right),
\end {equation}

\noindent and

\begin {equation} \label{eq10}
   N_{{\bm k}_2} = \frac{\displaystyle 2\pi\beta\hbar^2}{\displaystyle m_{\rm e}}~ \alpha n~ \exp \left(\!-\beta~\frac{\displaystyle \hbar^2k_2^2}{\displaystyle 2m_{\rm e}}\right).
\end {equation}

\noindent The coefficient $\xi$ in Eq.~(\ref{eq8}) ensures that there is no divergence when the photon energy $\hbar\Omega$ is equal to the gap energy $E_{\rm g}$. The exciton binding energy is $E_{\rm b}^{\rm X}$, so the total energy of an exciton is: $E_{\rm X}=E_{\rm g}-E_{\rm b}^{\rm X}+\hbar^2k_{\rm {cm}}^2/2M$.

To proceed with the calculations we combine Eqs.~(\ref{eq7}-\ref{eq10}), which yields:

\begin{eqnarray} \label{eq21}
&&R_{\rm {eX}}(\hbar\Omega) = \frac{\displaystyle 2\pi}{\displaystyle \hbar}\frac{\displaystyle 4\pi\xi \hbar\Omega /E_{\rm g}}{\displaystyle ( 1-\hbar^2\Omega^2 /E_{\rm g}^2)^2 +4\pi\xi} \sum_{{\bm k}_{\rm {cm}}} \frac{\displaystyle 2\pi\beta\hbar^2}{\displaystyle M}~ (1-\alpha)n~\exp \left(\!-\beta~ \frac{\displaystyle \hbar^2k_{\rm {cm}}^2}{\displaystyle 2M}\right)
\\
&&\times \sum_{{\bm k}_2} \frac{\displaystyle 2\pi\beta\hbar^2}{\displaystyle m_{\rm e}}~ \alpha n~ \exp \left(\!-\beta\hspace{0.05cm}\frac{\displaystyle \hbar^2k_2^2}{\displaystyle 2m_{\rm e}}\right)~\left|V_{\rm {scat}}({\bm k}_{\rm {cm}},{\bm k}_2)\right|^2~\delta \left (E_{\rm g}-E_{\rm b}^{\rm X}+\frac{\displaystyle \hbar^2k_{\rm {cm}}^2}{\displaystyle 2M}-\hbar\Omega- \frac{\displaystyle \hbar^2}{\displaystyle 2m_{\rm e}}(k_{\rm {cm}}^2+2{\bm k}_{\rm {cm}} \cdot {\bm k}_2)\right ),\nonumber
\end{eqnarray}

\noindent for the exciton-electron scattering. The function $R_{\rm {hX}}(\hbar\Omega)$ is formally identical to $R_{\rm {eX}}(\hbar\Omega)$ in Eq.~(\ref{eq21}). The effective masses $m_{\rm e}$ and $m_{\rm h}$ have only to be changed where appropriate. The rates $R_{\rm {cX}}(\hbar\Omega)$ depend explicitly on the degree of ionization, $\alpha$, via the product $\alpha(1 - \alpha)$. This product is maximum for $\alpha = 1/2$, which means that for given thermodynamic conditions on temperature and density, $R_{\rm {cX}}(\hbar\Omega)$, as defined above, is greater when the plasma is composed of a mixture of excitons and free carriers in equal proportions. As mentioned in section II.B, the nature of the electron-hole plasma in ZnSe QWs is closer to a plasma with $\alpha = 1/2$ than it is in GaAs QWs, and one should expect a more important contribution of exciton-carrier scattering to PL in ZnSe QWs. Note that very high proportion of excitons do not guarantee a great contribution to $R_{\rm {cX}}(\hbar\Omega)$ either, as it would simply mean that there would not be enough free carriers to scatter with. The key point here is to have $\alpha$ close to 1/2.

For numerical purposes, the discrete sums are approximated by 2D integrals: $\sum_{\bm k} \longrightarrow {\mathcal A}/4\pi^2 \int {\rm d}{\bm k}$. To evaluate the effects of exchange on the emission rates induced by exciton--carrier scattering, we first calculate $R_{\rm {cX}}(\hbar\Omega)$, considering \emph{only} the direct term of the scattering matrix elements. Note that, as shown in Appendix B, the numerical calculation of $R_{\rm {cX}}(\hbar\Omega)$ is greatly simplified in this case since $V_{\rm {dir}}^{\rm {cX}}$ does not depend on ${\bm k}_2$ (see Eq.~(\ref{eqb9})). Inclusion of the exchange terms in the calculation of $R_{\rm {cX}}(\hbar\Omega)$ requires the full numerical computation of Eq.~(\ref{eq21}).

\section{Numerical results}

The materials parameters for GaAs-based quantum-wells are: $m_{\rm e}= 0.067 m_0$, $m_{\rm h}= 0.46 m_0$, where $m_0$ is the free electron mass, $\epsilon_{\rm r} = 12$, $E_{\rm g} = 1.45$ eV. For ZnSe we use: $m_{\rm e}= 0.15 m_0$, $m_{\rm h}= 0.60 m_0$, $\epsilon_{\rm r} = 8.8$, $E_{\rm g} = 2.7$ eV. As in Ref.~\cite{BEN69} we take $\xi\approx 10^{-3}$. The ionization degree is evaluated at $T = 300$ K for various plasma densities given in Table 1. To study $V_{\rm {scat}}^{\rm {cX}}$ as a function of the exciton center of mass wavevector, ${\bm k}_{\rm {cm}}$, we fix the modulus of the free carriers wavevectors, ${\bm k}_2$, evaluated from the mean velocity of the free carriers within the 300 K carrier gases. In GaAs QW, $k_2=2.32~a_B^{-1}$ and $k_2=6.07~a_B^{-1}$ for the electron and the hole respectively. In ZnSe QW, these values are $k_2=1.24~a_B^{-1}$ and $k_2=2.48~a_B^{-1}$. 

\begin{table*}
\caption{\label{tab:table1}Values of plasma screening parameter $q_{\rm s}$ scaled to the excitonic Bohr radius, and the corresponding densities $n$ in $10^{11} \mbox{cm}^{-2}$ and degrees of ionization $\alpha$, for ZnSe and GaAs QWs at $T$ = 300 K. The exciton binding energies, $E_{\rm b}^{\rm X}$, are given in excitonic Rydbergs.}
\begin{ruledtabular}
\begin{tabular}{cccccc}
 &\multicolumn{2}{c}{{\bf GaAs}}&\multicolumn{2}{c}{{\bf ZnSe}}\\
 $q_{\rm s} a_{\rm B}$ &$n $ &$\alpha (n)$ &$n $ &$\alpha (n)$ &$E_{\rm b}^{\rm X}$  \\ \hline
 0.10 &0.14 &0.93 &0.66 &0.41 &3.30\\
 0.32 &0.48 &0.86 &2.51 &0.34 &2.51\\
 1.0  &1.62 &0.78 &6.78 &0.39 &1.35\\
\end{tabular}
\end{ruledtabular}
\end{table*}

\subsection{The scattering matrix elements}

\begin {figure}[!rh]
\centering
\scalebox{.33}{\rotatebox{0}{\includegraphics*{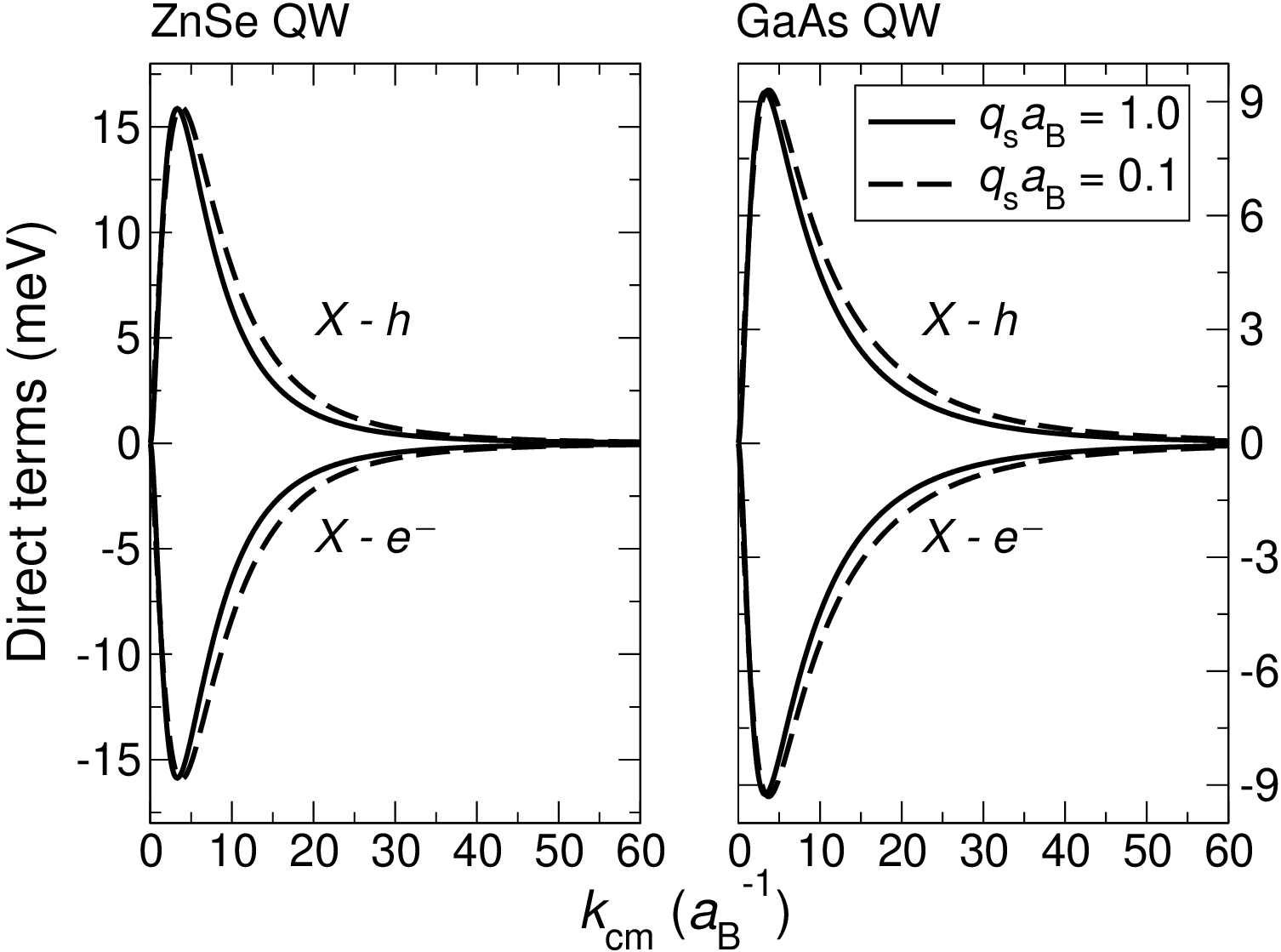}}}
\caption{Direct terms of the scattering matrix elements as functions of $k_{\rm cm}$, describing the 1$s$ exciton-carrier direct scattering within a screened Coulomb potential in ZnSe and GaAs quantum wells, in a low density regime at $T = 300$ K. For a given plasma screening $V_{\rm dir}^{\rm hX}$ is the exact opposite of $V_{\rm dir}^{\rm eX}$.}\label{fig2}
\end {figure}

To calculate the direct and exchange terms of the scattering matrix elements, Eqs.~(\ref{eq20}) and (\ref{eq20b}), we need to compute first the relative motion part of the exciton wavefunction in the screened Coulomb potential from the numerical solution of Eqs.~(\ref{eq13}) and (\ref{eq15}) obtained with the variable phase approach. In Fig.~\ref{fig2} the behavior of the direct terms of the scattering matrix elements versus $k_{\rm {cm}}$ is shown for  two values of the plasma screening: $q_{\rm s}a_{\rm B}$ = 0.1 and $q_{\rm s}a_{\rm B}$ = 1, for both materials. In all cases the qualitative behavior of $V_{\rm dir}^{\rm eX}$ is the same: $V_{\rm dir}^{\rm eX} = 0$ for $k_{\rm cm} = 0$ since the exciton cannot scatter from $k_{\rm {cm}} = 0$ to $k_{\rm {cm}} = 0$; then, for small wavenumbers, $V_{\rm dir}^{\rm eX}$ which is negative for all values of $k_{\rm {cm}}$, decreases and reaches a minimum for $k_{\rm {cm}}\approx 3a_{\rm B}^{-1}$, which means that the location of the minimum mostly depends on the ratio of the electron and hole effective masses, and less on the screening parameter in the low density regime. For exciton center of mass momenta larger than 3 $a_{\rm B}^{-1}$, $V_{\rm scat}^{\rm dir,e}$ is a monotonically increasing function of $k_{\rm {cm}}$ whose amplitude thus diminishes: the transfer of increasing large exciton center of mass momentum to its scattering partner is less likely. The matrix element $V_{\rm dir}^{\rm eX}$ is, as expected, greater for low screening, but not in a dramatic way: in the low density regime the screening parameter remains small enough not to have a siginificant impact on the amplitude of $V_{\rm dir}^{\rm eX}$. The magnitude of $V_{\rm dir}^{\rm eX}$ is also higher in ZnSe than it is in GaAs, reflecting the greater strength of Coulomb interaction in ZnSe. Note that if the electron and hole effective masses were equal, the electron-electron and electron-hole contributions to $V_{\rm dir}^{\rm eX}$ would cancel exactly and the scattering amplitude would be zero for all $k_{\rm {cm}}$. For a given plasma screening, the same comments apply to $V_{\rm dir}^{\rm hX}$, except that because of opposite signs, it is positive for all $k_{\rm cm}$. The functions $V_{\rm dir}^{\rm hX}$ and $V_{\rm dir}^{\rm eX}$ reach their extrema for the same values of $k_{\rm cm}$. Note that the extrema of $V_{\rm dir}^{\rm cX}$, ${\rm c} \equiv {\rm e},{\rm h}$, in both ZnSe-based and GaAs-based QWs, are reached for values of $k_{\rm {cm}}$ that are very close: here the material dependent effective masses have a limited influence on the behavior of the function $V_{\rm dir}^{\rm cX}(k_{\rm {cm}})$.

\begin {figure}[!rh]
\centering
\scalebox{.33}{\rotatebox{0}{\includegraphics*{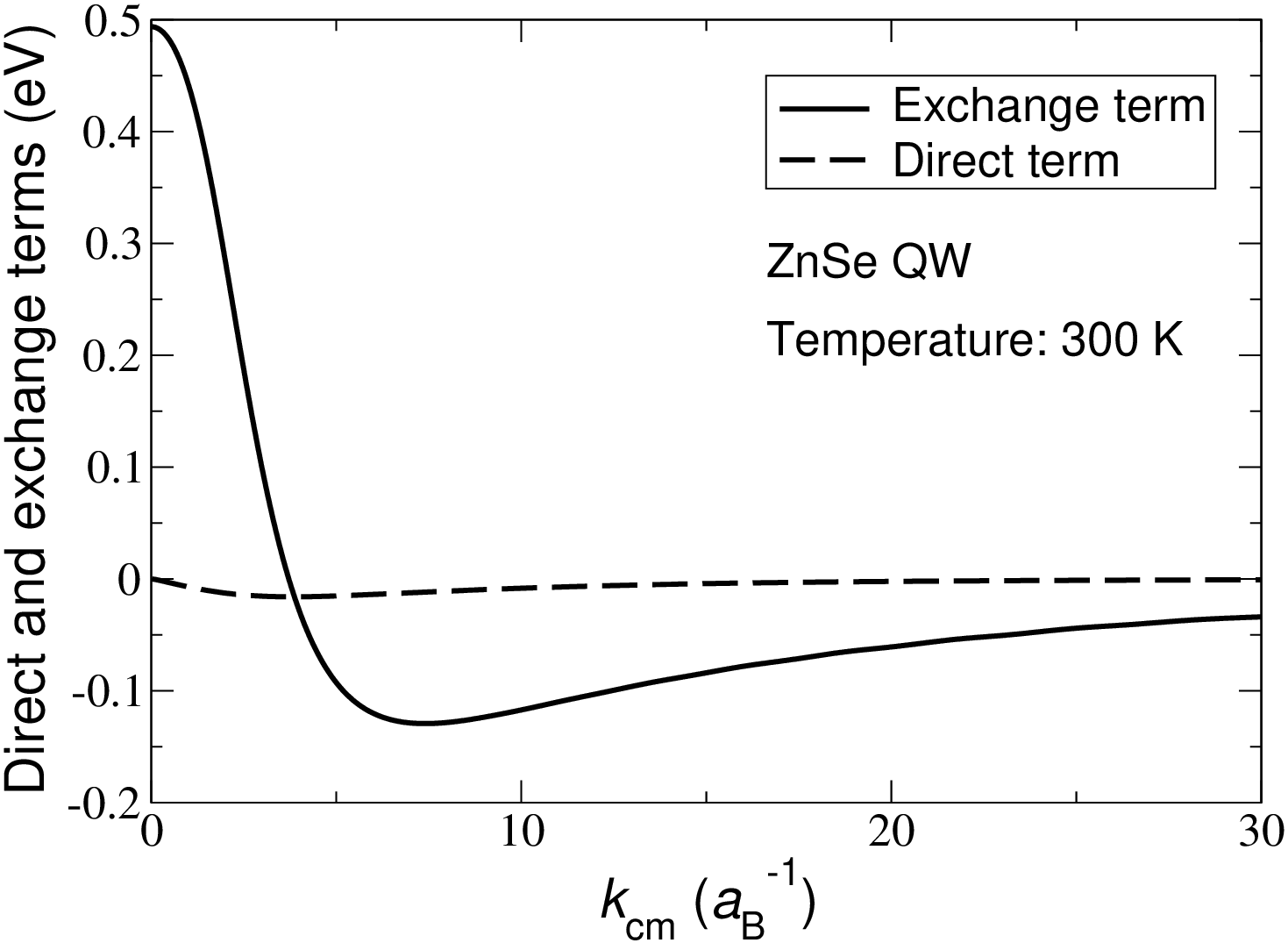}}}
\caption{Comparison of the direct and exchange terms of the $X-e^-$ scattering matrix elements as functions of $k_{\rm cm}$ in ZnSe for $q_{\rm s}a_{\rm B}=0.1$.}\label{fig3}
\end {figure}

We turn now to the study of the full scattering matrix elements,$V_{\rm {scat}}^{\rm {cX}}$, including the exchange terms. For discussion, the value of ${\bm k}_2$ is taken as the thermally averaged wavevector as introduced at the beginning of this section. A comparison of amplitudes of $V_{\rm dir}^{\rm eX}$ and $V_{\rm exch}^{\rm eX}$ is depicted in Fig.~\ref{fig3}. For a given value of $k_2$, $V_{\rm exch}^{\rm eX}$ is much greater than $V_{\rm dir}^{\rm eX}$. Exchange effects are therefore of importance even if the plasma is at room temperature. Another interesting observation is that unlike the direct term, the exchange term changes sign, thus reflecting the complex interplay of the Coulomb interactions between the three scattering partners. These results are consistent with the findings of Ramon \emph{et al} \cite{RAM03} in the case of electron--exciton scattering. Note that the calculations of Ramon \emph{et al} \cite{RAM03} were performed in the low temperature regime thus resulting in a greater amplitude of the exchange term at $k_{\rm cm} \approx 0$.

\begin {figure}[!rh]
\centering
\scalebox{.33}{\rotatebox{0}{\includegraphics*{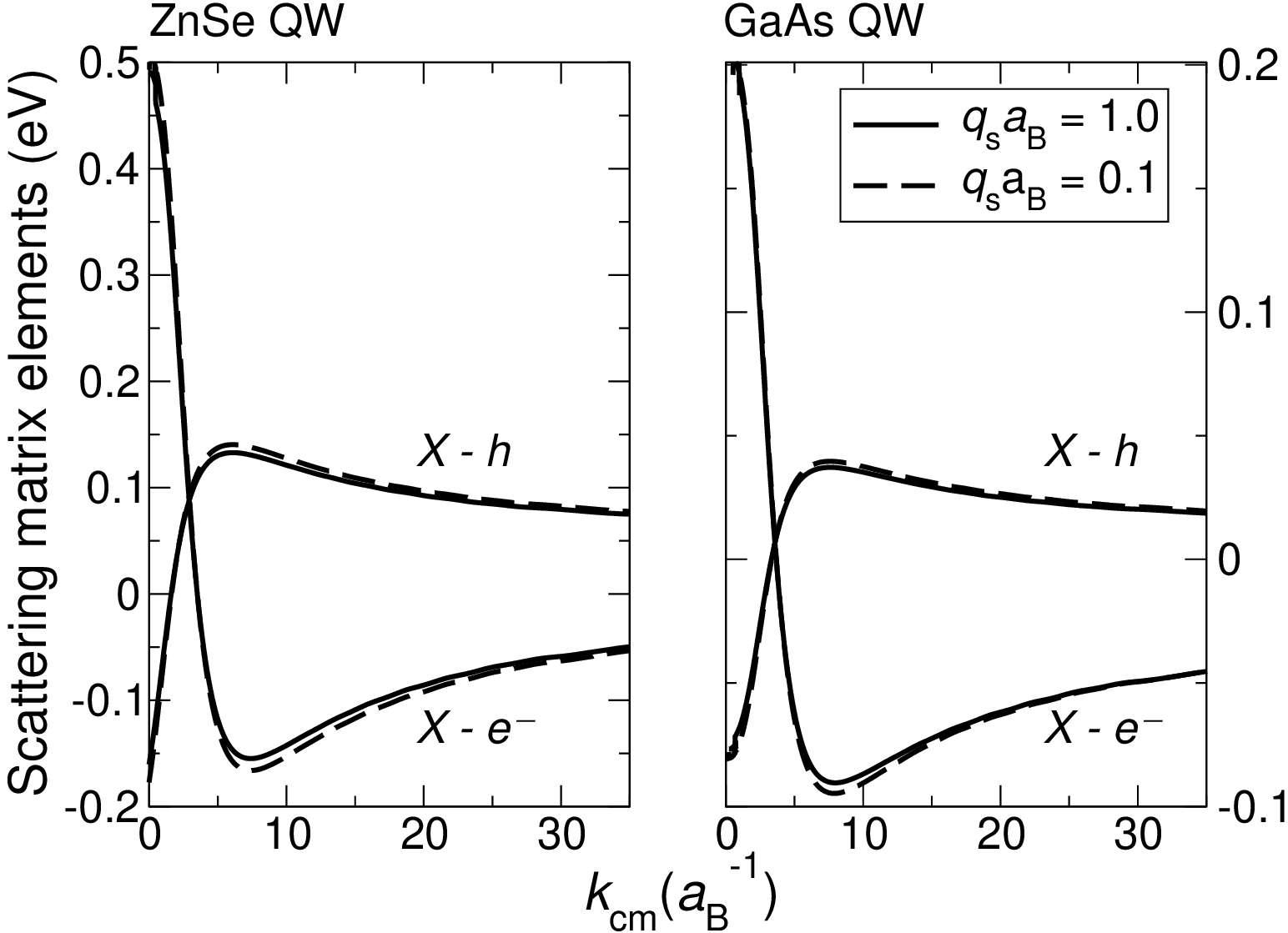}}}
\caption{Scattering matrix elements (approximated by the exchange terms) as functions of $k_{\rm cm}$, describing the 1$s$ exciton--carrier scattering within a screened Coulomb potential in ZnSe and GaAs quantum wells, in a low density regime at $T = 300$ K.}\label{fig4}
\end {figure}

The total scattering matrix elements can thus be safely approximated by the exchange terms as shown in Fig.~\ref{fig4}. For a given plasma screening, we find that the scattering matrix elements, $V_{\rm scat}^{\rm eX}$ and $V_{\rm scat}^{\rm hX}$, are not the exact opposite of each other, unlike the direct terms. This originates from the non-trivial dependence of $V_{\rm exch}^{\rm cX}$ on the free carriers wavevectors ${\bm k}_2$ and the effective masses as can be seen in Eq.~(\ref{eq20b}). For small and moderate exciton transferred momenta $k_{\rm cm}$, the amplitudes of the scattering matrix elements $V_{\rm scat}^{\rm hX}$ are smaller than those of $V_{\rm scat}^{\rm eX}$ simply because the exchange effects are less important with larger effective mass. However, as the value of $k_{\rm cm}$ increases the amplitudes become comparable and we observe that they decrease in a slower fashion for $V_{\rm scat}^{\rm hX}$ than for $V_{\rm scat}^{\rm eX}$. Because of their smaller effective mass, the electrons acquire a larger kinetic energy for a given value of the transferred momentum than the holes, which as a consequence diminishes the exchange effects for the lighter quasiparticles. The interplay between effective mass and transferred momentum thus has a non-trivial influence on the exchange term of the scattering matrix elements.

\subsection{Contribution of exciton--carrier scattering to the emission rates}

\begin {figure}[!rh]
\centering
\scalebox{.33}{\rotatebox{0}{\includegraphics*{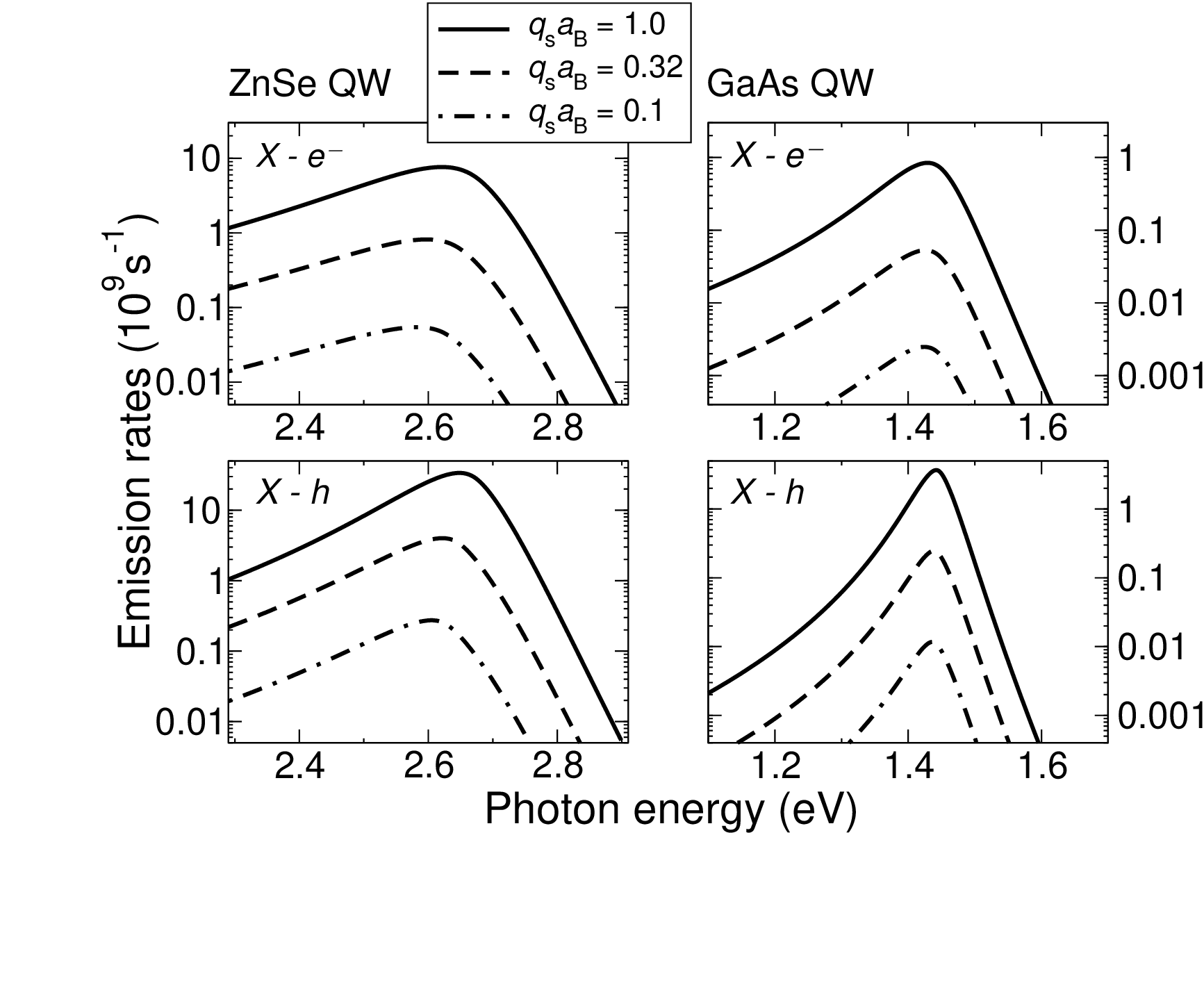}}}
\caption{Contribution of the direct 1$s$ exciton-carrier scattering to the emission rates evaluated for ZnSe and GaAs QWs in the nondegenerate regime at $T = 300$ K. The corresponding values of the densities for the screening parameters are given in Table~\ref{tab:table1}. The rates of photon emission induced by the direct exciton--hole scattering dominate those induced by the direct exciton--electron scattering in both materials, by one order of magnitude. Note the differing vertical scales.}\label{fig5}
\end {figure}

The numerical evaluation of Eq.~(\ref{eqb9}) gives the 1$s$ exciton-carrier \emph{direct} scattering contributions to the emission rates, shown in Fig.~\ref{fig5}. For a given material the behavior of $R_{\rm {cX}}(\hbar\Omega)$ reflects its complex density dependence via the factor $\alpha(n)(1-\alpha(n))n^2$. Note that since the scattering matrix element does not depend dramatically on $n$ it has a rather small influence on the emission rates: $R_{\rm {cX}}(\hbar\Omega)$ is proportional to the square of $V_{\rm dir}^{\rm cX}$ and while the amplitudes of $V_{\rm dir}^{\rm cX}$ decrease as the plasma density increases, the function $R_{\rm {cX}}(\hbar\Omega)$ is increasing significantly. It is also of interest to note that the magnitude of $R_{\rm {hX}}(\hbar\Omega)$, induced by the \emph{direct} term of the scattering matrix elements, is greater than $R_{\rm {eX}}(\hbar\Omega)$ for comparable densities. This can be explained as follows: first, since the direct terms of the scattering matrix elements $V_{\rm dir}^{\rm hX}$ and $V_{\rm dir}^{\rm eX}$ have exactly the same magnitudes for a given carrier density, the $X-e^-$ direct scattering channel is not favored despite the electron lighter effective mass; however, as can be seen from Eq.~(\ref{eqb9}), $R_{\rm {cX}}(\hbar\Omega)$ increases with increased carrier effective mass $m_{\rm c}$ (which can be found in the prefactor before the integral). Second, we checked that the function of photon energy, $\hbar\Omega$, resulting from the computation of the integral in Eq.~(\ref{eqb9}) (without the prefactors) is greater in the case of $X-h$ direct scattering near its maximum. This mass effect can be interpreted in terms of the two-dimensional density of states that is indeed greater for the hole gas than it is for the electron gas. Note that the emission spectra exhibit a thermal tail on the low energy side of the spectrum mirroring the free carriers distribution functions.

\begin {figure}[!rh]
\centering
\scalebox{.33}{\rotatebox{0}{\includegraphics*{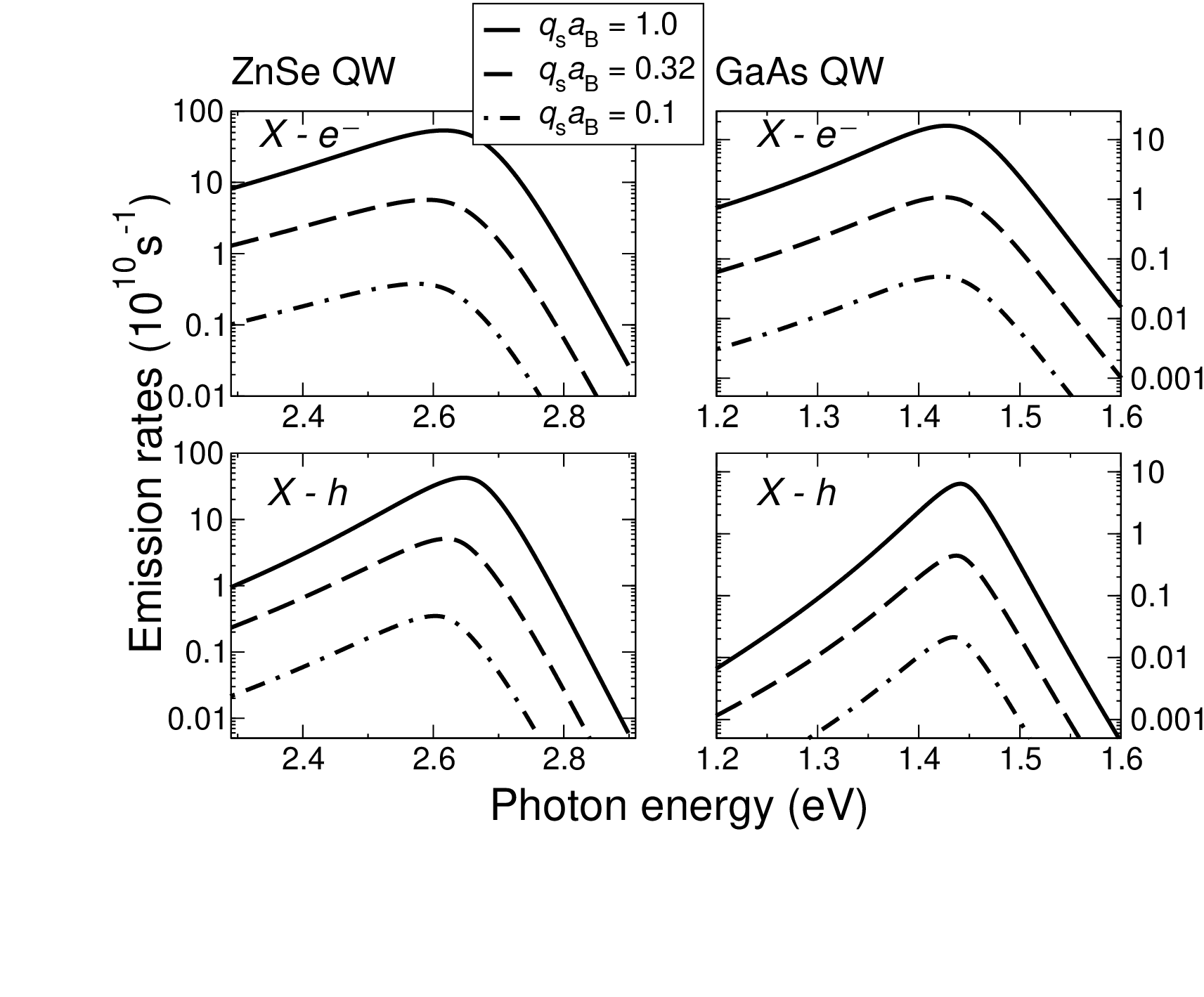}}}
\caption{Contribution of the 1$s$ exciton-carrier scattering to the emission rates evaluated for ZnSe and GaAs QWs in the nondegenerate regime at $T = 300$ K. The corresponding values of the densities for the screening parameters are given in Table~\ref{tab:table1}. Exchange has reversed the previously observed trend: it is now exciton--electron scattering that yields the highest photon emission rates in both materials. Note the differing vertical scales.}\label{fig6}
\end {figure}

To calculate the emission rates, including exchange effects in the scattering process, from Eq.~(\ref{eq21}), we should, in principle, take the carrier spins into account and hence study both the singlet and triplet configurations. However, as clearly shown in Fig.~\ref{fig4}, the direct term is negligible compared to the exchange term. Therefore, the total scattering matrix elements can be very well approximated by the exchange term only in Eq.~(\ref{eq19}), and the spin configuration becomes irrelevant here, as observed by Ramon \emph{et al} \cite{RAM03}. In Fig.~\ref{fig6}, we find that the inclusion of exchange effects in our model reverses the trend that we observed above: the magnitude of $R_{\rm {eX}}(\hbar\Omega)$ has increased dramatically and is now greater than that of $R_{\rm {hX}}(\hbar\Omega)$. Exchange is therefore the mechanism that makes the $X-e^-$ scattering a more efficient process than $X-h$ scattering to lead an exciton to the radiative cone where photon emission occurs. Although of importance, exchange effects are indeed reduced for the holes, which are heavier (4 times and 7 times the electron effective mass in ZnSe and GaAs respectively). Comparing results obtained for ZnSe and GaAs, it appears clearly that, for a given plasma screening, the contribution of exciton--carrier scattering to the emission rates at room temperature is more important in ZnSe than it is in GaAs, whatever the scattering channel is, $X-e^-$ or $X-h$. This is first due to the fact that for a given plasma screening, the plasma density is higher in ZnSe QW than it is in GaAs QW; second, because of a smaller dielectric constant that yields a strong exciton binding energy in ZnSe QWs, the exciton population remain important at low and moderate carrier densities (more than 50 \%) whereas the fraction of excitons in GaAs QWs remains modest (less than 20 \%) and hence yield a lower carrier-assisted excitonic contribution to photon emission. The nature of the mixed exciton/electron-hole plasma is therefore of importance.

\section{Summary and conclusion}

We have computed the exciton--electron and exciton--hole scattering matrix elements and discussed their properties. We have considered the fermionic nature of the carriers through exchange effects, which have been found to be very important even in a dilute mixed exciton/electron-hole plasma at room temperature. In these thermodynamic conditions the carrier-assisted radiative exciton recombination process has been investigated for GaAs-based and ZnSe-based quantum wells. For a given plasma screening, the emission rates due to elastic exciton--carrier scattering in ZnSe-based QWs are found to be one to two orders of magnitude greater than those in GaAs QWs. This is partly due to the fact that for the same value of plasma screening, the plasma density is greater in ZnSe QWs than it is in GaAs QWs, but more importantly, because of the lower ionization degree of the ZnSe QWs plasma. We have also found that in both systems the $X-h$ scattering channel is a more efficient one in the case of \emph{direct} scattering when the densities of electrons and holes are comparable. We attribute this essentially to the two-dimensional density of states which is greater for hole than for electron systems. The direct terms of the scattering matrix elements have been found to be negligible compared to the exchange terms. As a consequence, fermion exchange favors the $X-e^-$ scattering channel for the carrier-assisted radiative recombination.

\begin{acknowledgements}
We acknowledge support from the UK EPSRC. H. O. also acknowledges partial support from the Agence Nationale de la Recherche and from Science Foundation Ireland during his stay at the Tyndall National Institute, Cork, Ireland, where part of this work was done. H. O. is also pleased to thank Dr. Patrick Bogdanski for many fruitful discussions. M.E.P. acknowledges the hospitality of the ICCMP's staff and the financial support received from MCT and FINEP (Brazil).
\end{acknowledgements}

\appendix{

\section{Calculation of the scattering matrix elements}

In this appendix we show how the two terms of the scattering matrix elements whose expression is given below, are calculated in the case of exciton$-$electron scattering. We write $V_{\rm {scat}}^{\rm {eX}}({\bm k}_{\rm {cm}},{\bm k}_2)$ as follows:

\begin{eqnarray} \label{eq18}
V_{\rm {scat}}^{\rm {eX}}({\bm k}_{\rm {cm}},{\bm k}_2) & = & \frac{\displaystyle 1}{\displaystyle 2}\int {\rm d}{\bm r}_1{\rm d}{\bm r}_2{\rm d}{\bm r}_{\rm h}~\left(\phi_{{\bm k}_2+{\bm k}_{\rm {cm}}}^{\dagger}({\bm r}_2)\Psi_{0,\kappa,\vec{0}}^{\dagger} (\vec{0},\rho_{1{\rm h}}) - \phi_{{\bm k}_1+{\bm k}_{\rm {cm}}}^{\dagger}({\bm r}_1)\Psi_{0,\kappa,\vec{0}}^{\dagger} (\vec{0},\rho_{2{\rm h}})\right)
\\
&& \times \left(V_{\rm s}(\rho_{1{\rm h}})+V_{\rm s}(\rho_{2{\rm h}})+V_{\rm s}(\rho_{12})\right) ~\left(\Psi_{0,\kappa,{\bm k}_{\rm {cm}}}({\bm R}_1,\rho_{1{\rm h}})\phi_{{\bm k}_2}({\bm r}_2) -\Psi_{0,\kappa,{\bm k}_{\rm {cm}}}({\bm R}_2,\rho_{2{\rm h}})\phi_{{\bm k}_1}({\bm r}_1) \right),\nonumber
\end{eqnarray}

\noindent where ${\bm R}_1$ and ${\bm R}_2$ are the exciton center of mass coordinates, ${\bm r}_1$ and ${\bm r}_2$ the electrons coordinates and ${\bm r}_{\rm h}$ the hole coordinate. The relative distances are defined as follows: $\rho_{1{\rm h}}=|{\bm r}_{\rm h}-{\bm r}_1|$, $\rho_{2{\rm h}}=|{\bm r}_{\rm h}-{\bm r}_2|$ and $\rho_{12}=| {\bm r}_1-{\bm r}_2|$. Note that the attractive electron-hole interaction potentials $V_{\rm s}(\rho_{1{\rm h}})$ and $V_{\rm s}(\rho_{2{\rm h}})$ are negative and the repulsive electron-electron  interaction potential $V_{\rm s}(\rho_{12})$ is positive. Since we consider 1$s$ exciton states only, the value of the projection of the angular momentum, $m$, is zero and we remove the explicit $\varphi$ dependence from the expression of $\Psi_{0,\kappa,{\bm k}_{\rm {cm}}}$. The initial state of the free electron is the plane wave $\phi_{\bm k}$ characterized by the wavevector ${\bm k}$ (= ${\bm k}_1$ or ${\bm k}_2$), and its final state the plane wave $\phi_{{\bm k}+{\bm k}_{\rm {cm}}}$ characterized by ${\bm k}+{\bm k}_{\rm {cm}}$, where ${\bm k}_{\rm {cm}}$ is the exciton center of mass momentum that has been transfered during the scattering process. Equation (\ref{eq18}) is similar to the expression in Ref.~\cite{KAV03}.

Starting from Eq.~(\ref{eq18}), the first step consists of identifying the relevant physical terms: the direct and exchange terms, $V_{\rm {dir}}^{\rm {eX}}$ and $V_{\rm {exch}}^{\rm {eX}}$ respectively. Calculations yield the following expressions:

\begin{equation} \label{eqa2}
V_{\rm {dir}}^{\rm {eX}} = \int \phi_{{\bm k}_2+{\bm k}_{\rm {cm}}}^{\dagger}({\bm r}_2)\Psi_{0,\kappa,\vec{0}}^{\dagger}(\vec{0},\rho_{1{\rm h}}) \left(V_{\rm s}(\rho_{2{\rm h}})+V_{\rm s}(\rho_{12})\right) \Psi_{0,\kappa,{\bm k}_{\rm {cm}}}({\bm R}_1,\rho_{1{\rm h}})\phi_{{\bm k}_2}({\bm r}_2)~{\rm d}{\bm r}_1{\rm d}{\bm r}_2{\rm d}{\bm r}_{\rm h}
\end{equation}

\noindent and

\begin{equation} \label{eqa3}
V_{\rm {exch}}^{\rm {eX}} = -\int \phi_{{\bm k}_2+{\bm k}_{\rm {cm}}}^{\dagger}({\bm r}_2) \Psi_{0,\kappa,\vec{0}}^{\dagger}(\vec{0},\rho_{1{\rm h}}) \left(V_{\rm s}(\rho_{1{\rm h}})+V_{\rm s}(\rho_{2{\rm h}})+V_{\rm s}(\rho_{12})\right) \Psi_{0,\kappa,{\bm k}_{\rm {cm}}}({\bm R}_2,\rho_{2{\rm h}})\phi_{{\bm k}_1}({\bm r}_1)~{\rm d}{\bm r}_1{\rm d}{\bm r}_2{\rm d}{\bm r}_{\rm h}.
\end{equation}

\noindent An additional term $V_{\rm {add}}$ appears:

\begin{equation} \label{eqa4}
V_{\rm {add}} = \int \phi_{{\bm k}_2+{\bm k}_{\rm {cm}}}^{\dagger}({\bm r}_2)\Psi_{0,\kappa,\vec{0}}^{\dagger}(\vec{0},\rho_{1{\rm h}})~V_{\rm s}(\rho_{1{\rm h}})~ \Psi_{0,\kappa,{\bm k}_{\rm {cm}}}({\bm R}_1,\rho_{1{\rm h}})\phi_{{\bm k}_2}({\bm r}_2)~{\rm d}{\bm r}_1{\rm d}{\bm r}_2{\rm d}{\bm r}_{\rm h},
\end{equation}

\noindent and before proceeding with the calculations of $V_{\rm {dir}}^{\rm {eX}}$ and $V_{\rm {exch}}^{\rm {eX}}$, we show that $V_{\rm {add}}=0$. Recalling the explicit expression of the exciton wavefunction, $\Psi_{m,\kappa,{\bm k}_{\rm {cm}}}$, in Section III, the additional term may be rewritten as:

\begin{equation} \label{eqa5}
V_{\rm {add}} = \frac{\displaystyle 1}{\displaystyle {\mathcal A}^2}~\int e^{i({\bm k}_{\rm {cm}}+{\bm k}_2)\cdot{\bm r}_2}R_{0,\kappa}^2(\rho_{1{\rm h}})~V_{\rm s}(\rho_{1{\rm h}})~e^{i{\bm k}_{\rm {cm}}\cdot{\bm R}_1}e^{-i{\bm k}_2\cdot{\bm r}_2}~{\rm d}{\bm r}_1{\rm d}{\bm r}_2{\rm d}{\bm r}_{\rm h}.
\end{equation}

\noindent We define the Fourier tranforms of the square of the relative motion part of the excitonic wavefunction, Eq.~(\ref{eq10}):

\begin{equation}\label{eqa6}
R_{0,\kappa}^2(\rho_{1{\rm h}}) = \sum_{{\bm q}_1}{\widetilde {R_0^2}}_{,\kappa}({\bm q}_1)~ e^{i{\bm q}_1\cdot ({\bm r}_{\rm h}-{\bm r}_1)} =  \frac{\displaystyle {\mathcal A}}{\displaystyle 4\pi^2}\int {\widetilde {R_0^2}}_{,\kappa}({\bm q}_1)~ e^{i{\bm q}_1\cdot ({\bm r}_{\rm h}-{\bm r}_1)}{\rm d}{\bm q}_1,
\end{equation}

\noindent and of the scattering potential $V_{\rm s}(\rho_-)$:

\begin{equation}\label{eqa7}
V_{\rm s}(\rho_{1{\rm h}}) = \sum_{\bm q} \widetilde{V}_s({\bm q})~ e^{i{\bm q}\cdot({\bm r}_{\rm h}-{\bm r}_1)} = \frac{\displaystyle {\mathcal A}}{\displaystyle 4\pi^2} \int \widetilde{V}_s({\bm q})~ e^{i{\bm q}\cdot({\bm r}_{\rm h}-{\bm r}_1)}{\rm d}{\bm q},
\end {equation}

\noindent which we include into Eq.~(\ref{eqa5}) to find:

\begin{equation} \label{eqa8}
V_{\rm {add}} = \frac{\displaystyle 1}{\displaystyle 16\pi^4} \iint e^{i{\bm k}_{\rm {cm}}\cdot{\bm r}_2}e^{-i(\gamma_{\rm e}{\bm k}_{\rm {cm}} + {\bm q}_1+ {\bm q})\cdot{\bm r}_1} e^{-i(\gamma_{\rm h}{\bm k}_{\rm {cm}} - {\bm q}_1- {\bm q})\cdot{\bm r}_{\rm h}} {\widetilde {R_0^2}}_{,\kappa}({\bm q}_1)~\widetilde{V}_s({\bm q})~{\rm d}{\bm r}_1{\rm d}{\bm r}_2{\rm d}{\bm r}_{\rm h}{\rm d}{\bm q}_1{\rm d}{\bm q},
\end{equation}

\noindent where $\gamma_{\rm e} = m_{\rm e}/M$ and $\gamma_{\rm h} = m_{\rm h}/M$. Now, considering the identity $\int e^{i{\bm q}\cdot{\bm r}}{\rm d}{\bm r} = 4\pi^2\delta({\bm q})$, where $\delta$ denotes the Dirac function, the above integral reads:

\begin{equation} \label{eqa9}
V_{\rm {add}} = \iint \delta(\gamma_{\rm e}{\bm k}_{\rm {cm}}+{\bm q}_1+{\bm q}) ~\delta(-\gamma_{\rm h}{\bm k}_{\rm {cm}}+{\bm q}_1+{\bm q}) e^{i{\bm k}_{\rm {cm}}\cdot{\bm r}_2}{\widetilde {R_0^2}}_{,\kappa}({\bm q}) \widetilde{V}_s({\bm q})~ {\rm d}{\bm r}_2{\rm d}{\bm q}_1{\rm d}{\bm q},
\end{equation}

\noindent and, since $\int \delta(a-x)\delta(b-x)f(x)~{\rm d}x = 0$ when $a \neq b$, we finally find that $V_{\rm {add}}=0$.

\subsection{The exchange term}

We turn now to the exchange term $V_{\rm {exch}}^{\rm {eX}}$ that we artificially decompose into three terms: $V_{\rm {exch}}^{\rm {eX}} = V_{\rm {exch,1}}^{\rm {eX}}+V_{\rm {exch,2}}^{\rm {eX}}+V_{\rm {exch,3}}^{\rm {eX}}$:

\begin{equation} \label{eqa10}
V_{\rm {exch,1}}^{\rm {eX}} = -\int \phi_{{\bm k}_2+{\bm k}_{\rm {cm}}}^{\dagger}({\bm r}_2)\Psi_{0,\kappa,\vec{0}}^{\dagger} (\vec{0},\rho_{1{\rm h}})~ V_{\rm s}(\rho_{1{\rm h}})~ \Psi_{0,\kappa,{\bm k}_{\rm {cm}}}({\bm R}_2,\rho_{2{\rm h}})\phi_{{\bm k}_1}({\bm r}_1)~{\rm d}{\bm r}_1{\rm d}{\bm r}_2{\rm d}{\bm r}_{\rm h},
\end{equation}

\begin{equation} \label{eqa11}
V_{\rm {exch,2}}^{\rm {eX}} = -\int \phi_{{\bm k}_2+{\bm k}_{\rm {cm}}}^{\dagger}({\bm r}_2)\Psi_{0,\kappa,\vec{0}}^{\dagger} (\vec{0},\rho_{1{\rm h}})~ V_{\rm s}(\rho_{2{\rm h}})~ \Psi_{0,\kappa,{\bm k}_{\rm {cm}}}({\bm R}_2,\rho_{2{\rm h}})\phi_{{\bm k}_1}({\bm r}_1)~{\rm d}{\bm r}_1{\rm d}{\bm r}_2{\rm d}{\bm r}_{\rm h},
\end{equation}

\noindent and

\begin{equation} \label{eqa12}
V_{\rm {exch,3}}^{\rm {eX}} = -\int \phi_{{\bm k}_2+{\bm k}_{\rm {cm}}}^{\dagger}({\bm r}_1)\Psi_{0,\kappa,\vec{0}}^{\dagger} (\vec{0},\rho_{1{\rm h}})~ V_{\rm s}(\rho_{12})~ \Psi_{0,\kappa,{\bm k}_{\rm {cm}}}({\bm R}_2,\rho_{2{\rm h}})\phi_{{\bm k}_1}({\bm r}_1)~{\rm d}{\bm r}_1{\rm d}{\bm r}_2{\rm d}{\bm r}_{\rm h}.
\end{equation}

\noindent In the same fashion as above we define the Fourier tranforms:

\begin{equation}\label{eqa13}
R_{0,\kappa}(\rho_{1{\rm h}}) = \sum_{{\bm q}_1}\widetilde{R}_{0,\kappa}({\bm q}_1)~ e^{i{\bm q}_1\cdot ({\bm r}_{\rm h}-{\bm r}_1)} = \frac{\displaystyle {\mathcal A}}{\displaystyle 4\pi^2} \int \!\widetilde{R}_{0,\kappa}({\bm q}_1)~ e^{i{\bm q}_1\cdot ({\bm r}_{\rm h}-{\bm r}_1)}{\rm d}{\bm q}_1,
\end{equation}

\begin{equation}\label{eqa13b}
R_{0,\kappa}(\rho_{2{\rm h}}) = \sum_{{\bm q}_2}\widetilde{R}_{0,\kappa}({\bm q}_2)~ e^{i{\bm q}_2\cdot ({\bm r}_{\rm h}-{\bm r}_2)} = \frac{\displaystyle {\mathcal A}}{\displaystyle 4\pi^2} \int \!\widetilde{R}_{0,\kappa}({\bm q}_2)~ e^{i{\bm q}_2\cdot ({\bm r}_{\rm h}-{\bm r}_2)}{\rm d}{\bm q}_2,
\end{equation}

\begin{equation}\label{eqa14a}
V_{\rm s}(\rho_{2{\rm h}}) = \sum_{\bm q} \widetilde{V}_s^-({\bm q})~ e^{i{\bm q}\cdot({\bm r}_{\rm h}-{\bm r}_2)} = \frac{\displaystyle {\mathcal A}}{\displaystyle 4\pi^2} \int \widetilde{V}_s^-({\bm q})~ e^{i{\bm q}\cdot({\bm r}_{\rm h}-{\bm r}_2)}{\rm d}{\bm q},
\end {equation}

\noindent and

\begin{equation}\label{eqa14b}
V_{\rm s}(\rho_{12}) = \sum_{\bm q}\widetilde{V}_s^+({\bm q})~ e^{i{\bm q}\cdot({\bm r}_2-{\bm r}_1)} = \frac{\displaystyle {\mathcal A}}{\displaystyle 4\pi^2} \int \widetilde{V}_s^+({\bm q})~ e^{i{\bm q}\cdot({\bm r}_2-{\bm r}_1)}~{\rm d}{\bm q},
\end {equation}

\noindent where the $+$ and $-$ signs denote the repulsive and attractive potentials. Inserting the relevant Fourier transforms of Eqs.~(\ref{eqa6}), (\ref{eqa7}), (\ref{eqa13}-\ref{eqa14b}) into Eqs.~(\ref{eqa10}), (\ref{eqa11}) and (\ref{eqa12}) and performing the same type of calculations as for $V_{\rm {scat}}^{\rm {add}}$, we find:

\begin{equation} \label{eqa15}
V_{\rm {exch,1}}^{\rm {eX}} = - {\mathcal A}\int \widetilde{R}_{0,\kappa}(-{\bm k}_2 - {\bm q})\widetilde{R}_{0,\kappa}({\bm k}_2+\gamma_{\rm h}{\bm k}_{\rm {cm}}+{\bm q}) \widetilde{V}_s({\bm q})~{\rm d}{\bm q},
\end{equation}

\begin{equation} \label{eqa16}
V_{\rm {exch,2}}^{\rm {eX}} = - {\mathcal A}\widetilde{R}_{0,\kappa}({\bm k}_2 +\gamma_{\rm h}{\bm k}_{\rm {cm}}) \int \widetilde{R}_{0,\kappa}(-{\bm k}_2 - {\bm q}) \widetilde{V}_s^-({\bm q})~{\rm d}{\bm q},
\end{equation}

\noindent and

\begin{equation} \label{eqa17}
V_{\rm {exch,3}}^{\rm {eX}} = - {\mathcal A}\widetilde{R}_{0,\kappa}({\bm k}_2) \int \widetilde{R}_{0,\kappa}({\bm k}_2+\gamma_{\rm h}{\bm k}_{\rm {cm}}-{\bm q}) \widetilde{V}_s^+({\bm q})~{\rm d}{\bm q}.
\end{equation}

\noindent The exchange term, $V_{\rm {exch}}^{\rm {eX}}$, depends on the properties of each scattering partner: exciton binding energy, kinetic energies, effective masses of the electrons, hole and exciton. To obtain the exchange term for the exciton-hole scattering, one simply needs to swap the effecfive masses $m_{\rm e}$ and $m_{\rm h}$ where appropriate and change the overall sign.

\subsection{Direct term}

Finally, we show the main steps of the calculation of the direct term starting from Eq.~(\ref{eqa2}). We write $V_{\rm {dir}}^{{\rm eX}}=V_{\rm {dir}}^++V_{\rm {dir}}^-$ and we need only to explicit calculations for $V_{\rm {dir}}^-$ as $V_{\rm {dir}}^+$ has a similar structure. $V_{\rm {dir}}^-$ is the first of the two terms in Eq.~(\ref{eqa2}):

\begin{equation} \label{eqa18}
V_{\rm {dir}}^- = \int \phi_{{\bm k}_2+{\bm k}_{\rm {cm}}}^{\dagger}({\bm r}_2) \Psi_{0,\kappa,\vec{0}}^{\dagger}(\vec{0},\rho_{1{\rm h}}) V_{\rm s}(\rho_{2{\rm h}}) \Psi_{0,\kappa,{\bm k}_{\rm {cm}}}({\bm R}_1,\rho_{1{\rm h}})\phi_{{\bm k}_2}({\bm r}_2)~{\rm d}{\bm r}_1{\rm d}{\bm r}_2{\rm d}{\bm r}_{\rm h}
\end{equation}

\noindent Using the explicit expressions of the electron plane wave and exciton wavefunction of Section III as well as the definitions of the Fourier tranforms of Eqs.~(\ref{eqa6}) and (\ref{eqa14a}), we find:

\begin {equation} \label{eqa19}
V_{\rm {dir}}^- = \frac{\displaystyle 1}{\displaystyle 16\pi^4}\int e^{i({\bm k}_{\rm {cm}}+{\bm q})\cdot{\bm r}_2} ~e^{i({\bm q}_1 - \gamma_{\rm e}{\bm k}_{\rm {cm}})\cdot{\bm r}_1} ~e^{-i({\bm q}+{\bm q}_1 + \gamma_{\rm h}{\bm k}_{\rm {cm}})\cdot{\bm r}_{\rm h}}~{\widetilde {R_0^2}}_{,\kappa}({\bm q})\widetilde{V}_s^-({\bm q})~{\rm d}{\bm r}_1{\rm d}{\bm r}_2{\rm d}{\bm r}_{\rm h}{\rm d}{\bm q}{\rm d}{\bm q}_1.
\end {equation}

\noindent This 10-dimensional integral can be simplified using again the definition the $\delta$-function given above:

\begin{equation} \label{eqa20}
V_{\rm {dir}}^- = {\mathcal A} {\widetilde {R_0^2}}_{,\kappa}(\gamma_{\rm e}{\bm k}_{\rm {cm}}) \widetilde{V}_s^-({\bm k}_{\rm {cm}}).
\end{equation}

Doing the same calculations with $V_{\rm {scat}}^+$, Eq.~(\ref{eqa2}) can be reduced to the following expression in the Fourier space:

\begin {equation} \label{eqa21}
V_{\rm {dir}}^{\rm {eX}}({\bm k}_{\rm {cm}},\kappa) = {\mathcal A}\left({\widetilde {R_0^2}}_{,\kappa}(\gamma_{\rm e}{\bm k}_{\rm {cm}}) \widetilde{V}_s^-({\bm k}_{\rm {cm}})+{\widetilde {R_0^2}}_{,\kappa}(-\gamma_{\rm h}{\bm k}_{\rm {cm}}) \widetilde{V}_s^+({\bm k}_{\rm {cm}})\right).
\end {equation}

\noindent Since the Fourier transforms of the interaction potential and the squared wavefunctions are radial functions, $V_{\rm {dir}}^{\rm {eX}}({\bm k}_{\rm {cm}})$ only depends on the modulus of the vector ${\bm k}_{\rm {cm}}$. Combining Eq.~(\ref{eqa21}) with the Fourier transform of Eq.~(\ref{eq2}), we obtain the final expression of the direct term of the exciton$-$electron scattering matrix elements, which reads

\begin {equation} \label{eqa22}
V_{\rm {dir}}^{\rm {eX}}(k_{\rm {cm}},\kappa) = \frac{\displaystyle e^2}{\displaystyle 2\epsilon_0\epsilon_{\rm r}} \frac{\displaystyle {\widetilde {R_0^2}}_{,\kappa}(\gamma_{\rm h}k_{\rm {cm}}) - {\widetilde {R_0^2}}_{,\kappa}(\gamma_{\rm e}k_{\rm {cm}})}{\displaystyle k_{\rm {cm}}+q_{\rm s}}.
\end {equation}

\noindent Note that unlike the exchange term, the direct term only depends on the exciton properties: kinetic energy, binding energy and effective masses. We find that $V_{\rm {dir}}^{\rm {hX}}(k_{\rm {cm}})$ is the exact opposite of $V_{\rm {dir}}^{\rm {eX}}(k_{\rm {cm}})$. This is due to the fact that both free carriers wavefunctions are assumed to be plane waves.

\section{Contribution of the exciton-electron direct scattering to photon emission}

To calculate the contribution of the exciton-electron direct scattering to photon emission, $R_{\rm {eX}}$, we start from the evaluation of the discrete sum over all the vectors ${\bm k}_2$ in Eq.~(\ref{eq21}). In this case $V_{\rm {scat}}^{\rm {eX}}(k_{\rm {cm}},k_2)$ reduces to $V_{\rm {dir}}^{\rm {eX}}(k_{\rm {cm}})$ so that the scattering matrix element can be taken out of the sum over the vectors ${\bm k}_2$. To proceed, one can approximate the sum $S$:

\begin {equation}\label{eqb2}
S = \sum_{{\bm k}_2}\exp \left(\!-\beta~ \frac{\displaystyle \hbar^2k_2^2}{\displaystyle 2m_{\rm e}}\right)~ \delta \left(E_{\rm g}-E_{\rm b}^{\rm X}+\frac{\displaystyle \hbar^2k_{\rm {cm}}^2}{\displaystyle 2M}-\hbar\Omega- \frac{\displaystyle \hbar^2}{\displaystyle 2m_{\rm e}}(k_{\rm {cm}}^2+2{\bm k}_{\rm {cm}} \cdot {\bm k}_2)\right),
\end {equation}

\noindent by a 2D integral:

\begin {equation} \label{eqb3}
S = \frac{\displaystyle {\mathcal A}}{\displaystyle 4\pi^2}~ \int_{k_2^{\rm {min}}}^{k_2^{\rm {max}}}\!\int_0^{2\pi}\!\!k_2~\exp \left(\!-\beta\hspace{0.05cm}\frac{\displaystyle \hbar^2k_2^2}{\displaystyle 2m_{\rm e}}\right) \times \delta \left(E_{\rm g}-E_{\rm b}^{\rm X}+\frac{\displaystyle \hbar^2k_{\rm {cm}}^2}{\displaystyle 2M}-\hbar\Omega- \frac{\displaystyle \hbar^2}{\displaystyle 2m_{\rm e}}(k_{\rm {cm}}^2+2{\bm k}_{\rm {cm}} \cdot {\bm k}_2)\right) ~{\rm d}k_2{\rm d}\theta.
\end {equation}

The argument of the $\delta$ function in Eq.~(\ref{eqb3}), is of the form $X-Y\cos\theta$, where $\theta$ is the angle between the vectors ${\bm k}_{\rm cm}$ and ${\bm k}_2$. The integral over $\theta$ , $I\left[\delta\right]$, can hence be calculated analytically:

\begin {equation} \label{eqb4}
\int_0^{2\pi}\delta(X-Y\cos\theta){\rm d}\theta = \frac{\displaystyle 2}{\displaystyle |Y\sin[\cos^{-1}X/Y]|}
\end {equation}

\noindent if $|X|<|Y|$. We find:

\begin {equation} \label{eqb5}
I\left[\delta\right]  = \frac{\displaystyle 2}{\displaystyle \left| \frac{\displaystyle \hbar^2}{\displaystyle m_{\rm e}}k_{\rm {cm}}k_2\sin\left[\cos^{-1}\left(\frac{\displaystyle E_{\rm X} -\hbar\Omega-\hbar^2k_{\rm {cm}}^2/2m_{\rm e}}{\displaystyle \hbar^2k_{\rm {cm}}k_2/m_{\rm e}}\right)\right]\right|},
\end {equation}

\noindent where $X = E_{\rm X}-\hbar\Omega-\hbar^2k_{\rm {cm}}^2/2m_{\rm e}$, $Y = \hbar^2k_{\rm {cm}}k_2/m_{\rm e}$, and with the above condition on X and Y, the limits of the integral Eq.~(\ref{eqb3}) are: $k_2^{\rm {min}}=\left(E_{\rm X} -\hbar\Omega-\hbar^2k_{\rm {cm}}^2/2m_{\rm e}\right)m_{\rm e}/\hbar^2 k_{\rm {cm}}$ and $k_2^{\rm {max}}\rightarrow \infty$.

Considering the identity $\left|\sin[\cos^{-1}\Theta]\right| = \sqrt{1-\cos^2[\cos^{-1}\Theta]}=\sqrt{1-\Theta^2}$ and combining Eqs.~(\ref{eqb3}) and (\ref{eqb5}) lead to:

\begin {equation} \label{eqb6}
S=\frac{\displaystyle {\mathcal A}}{\displaystyle 2\pi^2}\int_{k_2^{\rm {min}}}^\infty\frac{\displaystyle k_2~\exp \left(\!-\beta~\frac{\displaystyle \hbar^2k_2^2}{\displaystyle 2m_{\rm e}}\right) {\rm d}k_2}{\displaystyle \sqrt{\frac{\displaystyle \hbar^4k_{\rm {cm}}^2k_2^2}{\displaystyle m_{\rm e}^2} - \left(E_{\rm X} - \hbar\Omega - \frac{\displaystyle \hbar^2k_{\rm {cm}}^2}{\displaystyle 2m_{\rm e}}\right)^2}}.
\end {equation}

\noindent Note that the analytical calculation would stop at this stage if the full scattering matrix element $V_{\rm {scat}}(k_{\rm {cm}},k_2)$ were considered, since it would explicitly appear in the above integral. In the case of \emph{direct} scattering, $V_{\rm {scat}}$ only depends on $k_{\rm {cm}}$ and hence we may proceed with the analytical calculations as follows.

With two successive changes of variables: $K_2=k_2^2$ and $K_2'=K_2/k_2^{\rm {min^2}}$, Eq.~(\ref{eqb6}) becomes:

\begin {equation} \label{eqb7}
	S=\frac{\displaystyle {\mathcal A}m_{\rm e}k_2^{\rm {min}}}{\displaystyle 4\pi^2\hbar^2k_{\rm {cm}}}~\int_1^\infty \frac{\displaystyle \exp \left(\!-\beta\hspace{0.05cm}\frac{\displaystyle \hbar^2k_2^{\rm {min^2}}}{\displaystyle 2m_{\rm e}}K_2'\right)}{\displaystyle \left(K_2' -1\right)^{1/2}}~{\rm d}K_2'.
\end {equation}

Considering the identity:$ \int_1^\infty e^{-\mu x} (x-1)^{-1/2}~{\rm d}x=\sqrt{\pi/\mu}~e^{-\mu}$, we find:

\begin{equation} \label{eqb8}
	S = \frac{\displaystyle {\mathcal A}m_{\rm e}}{\displaystyle 4\pi^2\hbar^2k_{\rm {cm}}}~\left(\frac{\displaystyle 2\pi m_{\rm e}}{\displaystyle \beta\hbar^2} \right)^{1/2} \exp \left(\!-\beta~ \frac{\displaystyle \hbar^2k_2^{\rm {min^2}}} {\displaystyle 2m_{\rm e}}\right).
\end{equation}

\noindent Finally, inserting Eq.~(\ref{eqb8}) into Eq.~(\ref{eq21}) and approximating the discrete sum over all the vectors ${\bm k}_{\rm {cm}}$ by an integral lead to the expression of the emission rate, $R_{\rm {eX}}(\hbar\Omega)$:

\begin {eqnarray}\label{eqb9}
&&R_{\rm {eX}}(\hbar\Omega) = \frac{\displaystyle \alpha(1-\alpha)N^2 \sqrt{2\pi m_{\rm e} \beta^3}}{\displaystyle M}\frac{\displaystyle 4\pi\xi \hbar\Omega /E_{\rm g}}{\displaystyle ( 1-\hbar^2\Omega^2 /E_{\rm g}^2)^2 +4\pi\xi}\\
\nonumber
&& \times
\int_0^{\infty}\exp-\displaystyle \beta \left[ \frac{\displaystyle \hbar^2k_{\rm {cm}}^2}{\displaystyle 2M}+\frac{\displaystyle m_{\rm e}}{\displaystyle 2\hbar^2k_{\rm {cm}}^2}\left(E_{\rm g}-E_{\rm b}^{\rm X}-\hbar\Omega-\frac{\displaystyle m_{\rm h}}{\displaystyle m_{\rm e}}\frac{\displaystyle \hbar^2k_{\rm {cm}}^2}{\displaystyle 2M}\right)^2\right]|V_{\rm {scat}}^{\rm {eX}}(k_{\rm {cm}})|^2{\rm d}k_{\rm {cm}}.
\end{eqnarray}

\noindent The expression of $R_{\rm {hX}}(\hbar\Omega)$ is simply obtained by swapping the effective masses $m_{\rm e}$ and $m_{\rm h}$ where appropriate.
}

\end{document}